\documentclass[10.75pt, a4paper]{article}
\usepackage{titlesec}
\titleformat{\section}
  {\bf\sffamily}
  {\thesection. }
  {5pt}
  {\MakeUppercase}
\renewcommand{\thesection}{\Roman{section}} 

\titleformat{\subsection}
  {\bf\sffamily}
  {\thesubsection. }
  {5pt}{}
  
\renewcommand{\thesubsection}{\Alph{subsection}}

\usepackage[affil-it]{authblk} 
\usepackage{etoolbox}
\usepackage{lmodern}

\usepackage{eurosym}

\usepackage[utf8]{inputenc}
\usepackage{geometry}
\usepackage[hidelinks]{hyperref}
\usepackage[citestyle=numeric,style=phys,biblabel=brackets,backend=bibtex,sorting=none,url=false,doi=false, natbib]{biblatex}
\bibliography{bibliography}
\DefineBibliographyStrings{english}{andothers={\itshape et\addabbrvspace al\adddot}}

\usepackage{csquotes}
\usepackage[]{authblk} 
\usepackage{etoolbox}
\usepackage{lmodern}

\usepackage{amsmath}
\usepackage{enumerate}
\usepackage{enumitem}
\usepackage{graphicx}
\usepackage{siunitx}
\usepackage{float}
\usepackage[]{parskip} 

\makeatletter
\patchcmd{\@maketitle}{\LARGE \@title}{\fontsize{16}{19.2}\selectfont\@title}{}{}
\makeatother

\usepackage[normalem]{ulem}

\usepackage{graphicx}
\usepackage{dcolumn}
\usepackage{bm}
\usepackage{tikz}
\usetikzlibrary{math}
\usetikzlibrary{shapes.geometric, arrows}
\usetikzlibrary{positioning}
\usetikzlibrary{shapes,arrows}
\usetikzlibrary{intersections}
\usepackage{csvsimple}
\usepackage{changepage}
\usepackage[tbtags]{mathtools}
\usepackage{siunitx}
\usepackage{csquotes}

\usepackage{float}
\usepackage{subfigure}
\usepackage[
	nonumberlist, 				
	acronym,      				
	nomain,						
	nopostdot,					
]{glossaries}
\usepackage{glossary-superragged}
\usepackage[hidelinks]{hyperref}
\usepackage{color, colortbl}

\usepackage{wrapfig}

\usepackage{pgfplots}
\usepackage{tikz}
\usetikzlibrary{arrows}
\usepackage{amsmath}
\pgfplotsset{compat=newest}
\usepgfplotslibrary{fillbetween}
\usetikzlibrary{calc}
\def\centerarc[#1](#2)(#3:#4:#5)
{ \draw[#1] ($(#2)+({#5*cos(#3)},{#5*sin(#3)})$) arc (#3:#4:#5); }

\usepackage{amssymb}
\let\vec\mathbf
\usepackage{nicefrac}

\usepackage{abstract}

\usepackage{multirow}
\usepackage{tabularx}
\usepackage{booktabs}
\usepackage{array}
\usepackage{longtable}
\newcolumntype{L}[1]{>{\raggedright\let\newline\\\arraybackslash\hspace{0pt}}m{#1}}
\newcolumntype{C}[1]{>{\centering\let\newline\\\arraybackslash\hspace{0pt}}m{#1}}
\newcolumntype{R}[1]{>{\raggedleft\let\newline\\\arraybackslash\hspace{0pt}}m{#1}}

\newacronym{3d}{3D}{three dimensional}
\newacronym{am}{AM}{additive manufacturing}
\newacronym{fdm}{FDM}{fused deposition modeling}
\newacronym{ism}{ISM}{in-space manufacturing}
\newacronym{iss}{ISS}{International Space Station}
\newacronym{fcb}{FCB}{Functional Cargo Block}
\newacronym{dem}{DEM}{discrete element method}
\newacronym{md}{MD}{molecular dynamics}
\newacronym{dc}{DC}{direct-current}
\newacronym[plural=PFCs,firstplural=parabolic flight campaigns (PFCs)]{pfc}{PFC}{Parabolic Flight Campaign}
\newacronym{fft}{FFT}{Fast Fourrier Transform}
\newacronym{cad}{CAD}{Computer Assisted Design}
\newacronym{ptfe}{PTFE}{polytetrafluoroethylene}
\newacronym{ps}{PS}{polystyrene}
\newacronym{nasa}{NASA}{National Aeronautics and Space Administration}
\newacronym{esamm}{ESAMM}{Extended Structure Additive Manufacturing Machine}
\newacronym{amf}{AMF}{Additive Manufacturing Facility}
\newacronym{us}{US}{United States}
\newacronym{usa}{USA}{United States of America}
\newacronym{bmgs}{BMGs}{Bulk Metallic Glasses}
\newacronym{esa}{ESA}{European Space Agency}
\newacronym{si}{SI}{International System of Units, abbreviated from French \textit{Syst\`{e}me International (d'unit\'{e}s)}}
\newacronym{dlr}{DLR}{German Aerospace Center}
\newacronym{liggghts}{LIGGGHTS}{\acrshort{lammps} Improved for General Granular and Granular Heat Transfer Simulations}
\newacronym{lammps}{LAMMPS}{Large-scale Atomic/Molecular Massively Parallel Simulator}
\newacronym{sjkr}{SJKR}{Simplified Johnson-Kendall-Roberts}
\newacronym{ded}{DED}{Directed Energy Deposition}
\newacronym{slm}{SLM}{Selective Laser Melting}
\newacronym{sls}{SLS}{Selective Laser Sintering}
\newacronym{eva}{EVA}{Extra-Vehicular Activity}
\newacronym{sem}{SEM}{Scanning Electron Microscopy}
\newacronym{RPM}{RPM}{Ramdom Positioning Machine}
\newacronym{rpm}{rpm}{revolutions per minute}
\newacronym{rise}{RISE}{Research Internships in Science and Engineering}
\newacronym{daad}{DAAD}{German Academic Exchange Service, abbreviated from German \textit{Deutscher Akademischer Austauschdienst}}
\newacronym{fsm}{FSM}{finite-state machine}
\newacronym{ir}{IR}{infrared}
\newacronym{pcbs}{PCBs}{Printed Circuit Boards}
\newacronym{pcb}{PCB}{Printed Circuit Board}
\newacronym{mcr}{MCR}{Modular Compact Rheometer}
\newacronym{sff}{SFF}{Solid Freeform Fabrication}
\newacronym{uv}{UV}{ultraviolet}
\newacronym{abs}{ABS}{acrylonitrile butadiene styrene}
\newacronym{hpde}{HPDE}{high density polyethylene}
\newacronym{pei}{PEI}{polyetherimide}
\newacronym{bff}{BFF}{BioFabrication Facility}
\newacronym{lens}{LENS}{Laser Engineered Net Shaping}
\newacronym{cnc}{CNC}{Computer Numerical Control}
\newacronym{ebf3}{EBF$^3$}{Electron Beam Free-Form Fabrication}
\newacronym{leo}{LEO}{Low Earth Orbit}
\newacronym{pc}{PC}{polycarbonate}
\newacronym{crissp}{CRISSP}{Customisable Recyclable International Space Station Packaging}
\newacronym{Athena}{Athena}{Advanced Telescope for High-ENergy Astrophysics}
\newacronym{lbm}{LBM}{Laser Beam Melting}
\newacronym{bam}{BAM}{Federal Institute for Materials Research and Testing, abbreviated from German \textit{Bundesanstalt f\"{u}r Materialforschung und-pr\"{u}fung}}
\newacronym{pbf}{PBF}{powder bed fusion}
\newacronym{eb}{EB}{Electron Beam}
\newacronym{2d}{2D}{two dimensional}
\newacronym{4d}{4D}{four dimensional}
\newacronym{ft4}{FT4}{Freeman Technology 4 Powder Rheometer}
\newacronym{dsc}{DSC}{Differential Scanning Calorimetry}
\newacronym{pmma}{PMMA}{polymethylmethacrylate}
\newacronym{1g}{$1g$}{gravity on-ground}
\newacronym{mug}{$\mu g$}{microgravity}
\newacronym{bcm}{BCM}{Box Counting Method}
\newacronym{mct}{MCT}{Mode Coupling Theory}
\newacronym{gmct}{gMCT}{granular Mode Coupling Theory}
\newacronym{itt}{ITT}{Integration Through Transients}
\newacronym{mfc}{MFC}{Mass Flow Controller}
\newacronym{ct}{CT}{computed tomography}
\newacronym{xct}{XCT}{X-ray computed tomography}
\newacronym{cv}{CV}{curriculum vitae}
\newacronym{pi}{PI}{principal investigator}
\newacronym{osp}{OSP}{orthogonal superimposed perturbation}
\newacronym{npi}{NPI}{Network Partnering Initiative}
\newacronym{ecsat}{ECSAT}{European Centre for Space Applications and Telecommunications}
\newacronym{eac}{EAC}{European Astronaut Centre}
\newacronym{estec}{ESTEC}{European Space Research and Technology Centre}
\newacronym{fps}{fps}{frames per second}
\newacronym{pdf}{pdf}{probability density function}
\newacronym{al}{Al}{aluminium}
\newacronym{ss}{\textit{SS}}{\textit{Smooth Surface}}
\newacronym{rs}{\textit{RS}}{\textit{Rough Surface}}
\newacronym{rcp}{rcp}{random close packing}
\newacronym{iop}{IoP UvA}{Institute of Physics of the University of Amsterdam}
\newacronym{mp}{MP}{Institute of Material Physics for Space}
\newacronym{elgra}{ELGRA}{European Low Gravity Research Association}
\newacronym{zarm}{ZARM}{Center of Applied Space Technology and Microgravity}
\newacronym{piv}{PIV}{particle image velocimetry}
\usepackage[framemethod=tikz]{mdframed}
\usepackage{lipsum}
\usepackage{dirtytalk}
\usepackage{tcolorbox}
 \usepackage{url}
 \usepackage{float}
\usepackage[noend]{algpseudocode}
\usepackage{algorithm, caption}
\usepackage[font=footnotesize]{caption}
\newtcolorbox{mybox}[1]{colback=green!6!white,colframe=black!75!black,fonttitle=\bfseries,title=#1}
\newtcolorbox{mybox2}{colback=red!5!white,colframe=red!75!black}

\usepackage{pifont}

\usepackage{soul,xcolor}
\setstcolor{red}

\usepackage{xcolor,hyperref}
\hypersetup{
   colorlinks,
   linkcolor={blue!50!black},
   citecolor={blue!50!black},
   urlcolor={blue!80!black}
} 

\definecolor{mycolor}{rgb}{0.122, 0.435, 0.698}

 \title{Billiards with Spatial Memory} 

\author
{Thijs Albers, Stijn Delnoij, Nico Schramma, Maziyar Jalaal\footnote{m.jalaal@uva.nl, ORCID: 0000-0002-5654-8505}$^{\ast}$}
\affil{Institute of Physics, University of Amsterdam; Science Park 904, Amsterdam, The Netherlands}

\begin{document}
\definecolor{brickred}{rgb}{0.8, 0.25, 0.33}
\definecolor{darkorange}{rgb}{1.0, 0.55, 0.0}
\definecolor{persiangreen}{rgb}{0.0, 0.65, 0.58}
\definecolor{persianindigo}{rgb}{0.2, 0.07, 0.48}
\definecolor{cadet}{rgb}{0.33, 0.41, 0.47}
\definecolor{turquoisegreen}{rgb}{0.63, 0.84, 0.71}
\definecolor{sandybrown}{rgb}{0.96, 0.64, 0.38}
\definecolor{blueblue}{rgb}{0.0, 0.2, 0.6}
\definecolor{ballblue}{rgb}{0.13, 0.67, 0.8}
\definecolor{greengreen}{rgb}{0.0, 0.5, 0.0}
\begingroup
\sffamily
\date{}
\maketitle
\endgroup

\begin{abstract}

Many classes of
active matter develop spatial memory by encoding information in space, leading to complex pattern formation. It has been proposed that spatial memory can lead to more efficient navigation and collective behaviour in biological systems and influence the fate of synthetic systems. This raises important questions about the fundamental properties of dynamical systems with spatial memory.
We present a framework based on mathematical billiards in which particles remember their past trajectories and react to them. Despite the simplicity of its fundamental deterministic rules, such a system is strongly non-ergodic and exhibits highly-intermittent statistics, manifesting in complex pattern formation. We show how these self-memory-induced complexities emerge from the temporal change of topology and the consequent chaos in the system. We study the fundamental properties of these billiards and particularly the long-time behaviour when the particles are self-trapped in an \textit{arrested} state. We exploit numerical simulations of several millions of particles to explore pattern formation and the corresponding statistics in polygonal billiards of different geometries. Our work illustrates how the dynamics of a single-body system can dramatically change when particles feature spatial memory and provide a scheme to further explore 
systems with complex memory kernels.


\textbf{keywords: Active Matter $|$ Memory $|$ Mathematical Billiard $|$ Chaos $|$ Pattern Formation} 

\end{abstract}


In cognitive psychology, spatial memory refers to the ability to remember and mentally map the physical spaces in the brain~\cite{tolman1948cognitive, o1978hippocampus, squire1992memory,eichenbaum1999hippocampus, mcnaughton2006path,o2014spatial}. It is an essential process in spatial awareness and to reach optimized navigation through complex environments, either for a taxi driver in London to find the fastest route~\cite{maguire2000navigation,woollett2009talent} or for a mouse to quickly find food in a maze~\cite{sharma2010assessment}. In fact, a variety of species with different levels of complexity, from honey bees~\cite{menzel2005honey,collett2013spatial}, and ants~\cite{heyman2019ants}, to birds~\cite{healy2004spatial}, bats~\cite{yartsev2013representation}, and human~\cite{tolman1948cognitive,burgess2006spatial} share this cognitive feature. 
Spatial memory, however, can also be achieved externally: in contrast to a cognitive map (where information is stored internally in the brain), the information is encoded in space itself, and then retrieved when the organism re-encounters it. Such memory 
can potentially enable collective behaviour in groups and optimize cost on the organismal level~\cite{cabanes2015ants,smith2017hansel}. External spatial memory is often mediated by chemical trails and, generally speaking, could be attractive (self-seeking) or repulsive (self-avoiding). 
Some species of bacteria are attracted to the bio-chemical trails they leave behind 
and by that, they form emergent complex patterns 
~\cite{budrene1991complex,budrene1995dynamics,mittal2003motility}. Examples of self-avoiding spatial memory can also be found in the slime mold \textit{Physarum polycephalum} — a eukaryotic multinucleated single cell — which forms spatial memory by leaving extracellular slime at the navigated location while searching for food. The slime then acts as a cue and the cell avoids those regions which have been explored already (also see~\cite{kramar2021encoding,bhattacharyya2022memory}).
Other biological examples can be found in epithelial cell migration when cells modify their external environment by reshaping their extracellular matrix or by secreting biochemical signalling cues~\cite{d2021cell,clark2022self}.

The self-avoiding spatial memory is not limited to living systems, but can also be observed in physico-chemically self-propelled particles that actively change the energy landscape 
in which they manoeuvre. An example is auto-phoretic active droplets which move due to interfacial stresses caused by surface tension gradients~\cite{thutupalli2011swarming,moerman2017solute,jin2017chemotaxis,hokmabad2021emergence,hokmabad2022chemotactic}. Active droplets leave a chemical trail behind as they move around and avoid these trails due to the local change in concentration gradients. Similar self-avoiding behaviour had been observed in other self-propelling active \say{particles} such as spider molecules~\cite{pei2006behavior,hamming2020influenza} and even 
nano-scale surface alloying islands~\cite{schmid2000alloying}.

Understanding and predicting the dynamics of active systems with memories is a difficult task. Most experimental systems are highly nonlinear and include probabilistic features that are often time and material dependent. Additionally, the interaction with the boundaries presents more complexities.
Here, we ask the question of how a dynamical system with self-avoiding memory behaves in two dimensions? We present a fully deterministic model with minimal ingredients for motile particles with spatial memory. We report that even such a simple single-body dynamical system exhibits chaos and complex interactions with boundaries, resulting in anomalous dynamics and surprisingly highly-intermittent behavior.


Consider a classical billiard: a mass-less point-particle moves ballistically on a closed two-dimensional domain $\Omega \subset \mathbb{R}^2$. The particle has a constant 
speed and does not experience any frictional/viscous dissipation. When reaching a boundary $\partial \Omega$, the particle follows an elastic reflection, \emph{i.e.}, the angle of incidence is equal to the angle of reflection. 
%
For over a century, mathematical billiards of various shapes have been studied by physicists and mathematicians to understand dynamical systems and geometries related to various problems, from the \emph{theory of heat and light}~\cite{kelvin1901nineteenth,sinai1973ergodic}, and (often Riemannian) surfaces~\cite{veech1989teichmuller, kerckhoff1986ergodicity,gutkin2012billiard,eskin2018invariant,wright2016rational,eskin2020billiards, gutkin1984billiards, mcmullen2003billiards, frkaczek2014non}, to chaos in classical, semi-classical and quantum systems~\cite{berry1981quantizing,heller1984bound,tomsovic1993long,nockel1997ray,dembowski2000first,stone2010chaotic,serbyn2021quantum,ravnik2021quantum,lozej2022quantum}.

\begin{figure}[h!]
    \centering
    \includegraphics[width=0.8\textwidth]{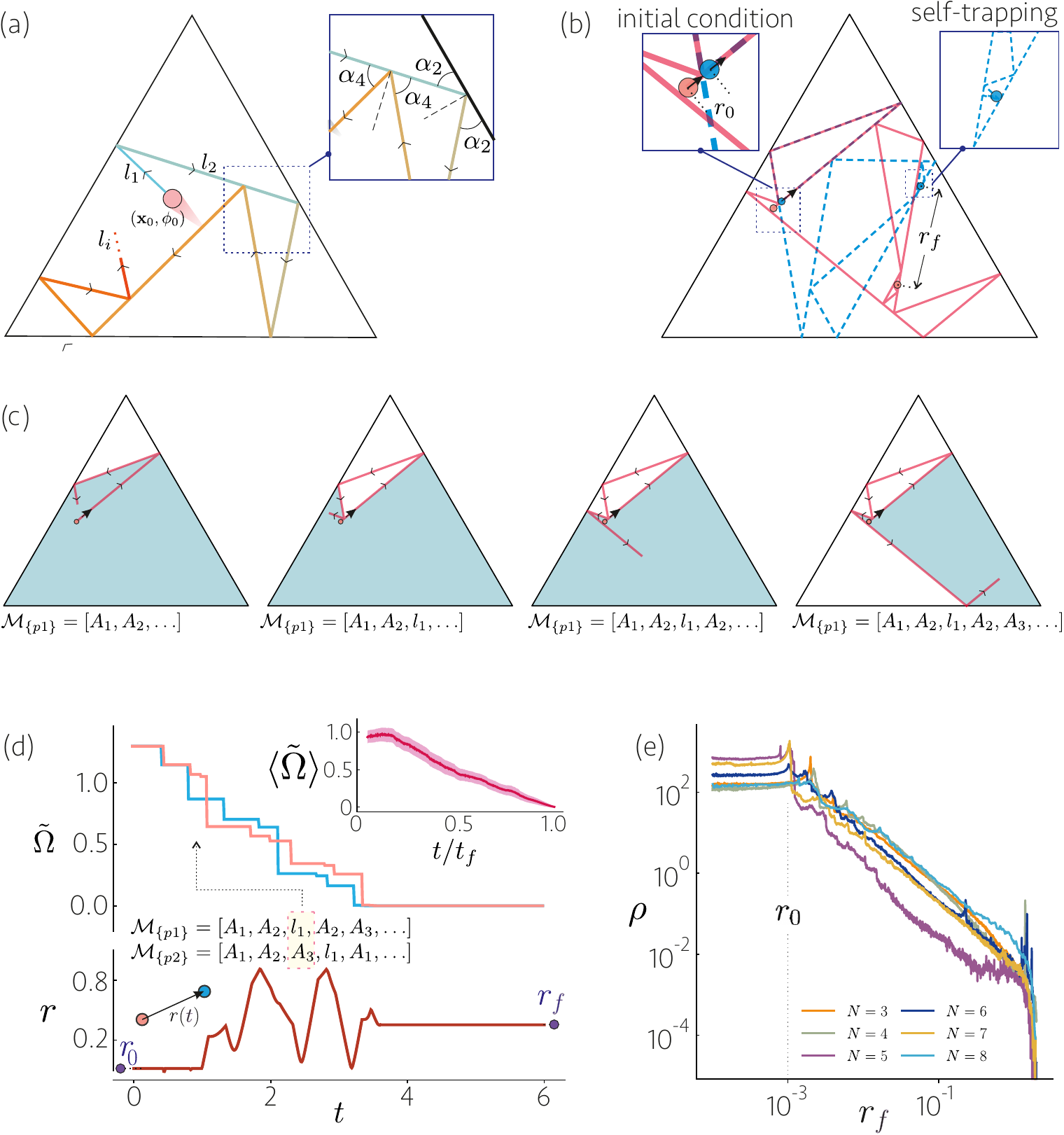}
    \caption{a) Underlying principle of Self-Avoiding Billiards: a particle moves ballistically from the initial condition $(\mathbf{x}_0, \phi_0)$ where $\mathbf{x}_0$ is the initial position vector and $\phi_0$ is the initial angle. The particle elastically reflects on the boundaries and its own trajectories, creating a line segment of length $l_i$, where $i\in [1,\infty)$ is the number of segments. The particle moves until it self-traps itself in a singular point, where the total length of the trajectory is $\mathcal{L}=\Sigma \, l_i$ (see video 1). 
    b) Two particles at initially close distance $r_0 = \vert \mathbf{x}^2_0 -\mathbf{x}^1_0 \vert$ and identical $\phi_0$ diverge significantly from their path and self-trap on distinct locations at a distance $r_f$ (see video 2).
    c) A self-avoiding particle changes the effective geometry of the billiard ($\Omega$, highlighted in blue) over time. $\mathcal{M}_{p_i}$ denotes the incident vector of the particle where $A_1, A_2,$ and $A_3$ are the right, left, and bottom edges of the triangle, respectively, and $l_1$ is the first segments of the trajectory.
    d) The dynamics of effective billiard area $\tilde{\Omega}$ and the distance between the two particles, $r$. The inset shows the ensemble average of the effective area \emph{vs.} time normalized by the trapping time $t_f$. 
    e) Distribution of $r_f$ for polygons with $\mathcal{N} \in [3,8]$. In all cases, $r_0=10^{-3}$ and the curves are averaged over $10^5$ samples.}
    \label{fig1}
\end{figure}

Here, we present a billiard with memory. 
In contrast to classical billiards, the particle continuously modifies the topology of the billiard table, creating spatial memory. We consider the simplest type of self-avoiding spatial memory: the particle reflects on its own trajectories from the past and avoids them in the same way it reflects on the boundaries (see figure~\ref{fig1}a and supplementary video 1). 
This Self-Avoiding Billiard (SAB) features a series of interesting properties. 
First, it fundamentally lacks periodic orbits (closed geodesics), as the particle cannot follow its past. 
Second, the particle presents a continuous-time dynamical system with self-induced excluded-volume (see figure~\ref{fig1}c). This means, in the long term, 
the particle reduces the size of its domain by consecutive intersections,
\emph{i.e.}, $\tilde{\Omega}(t) \rightarrow 0$, where $\tilde{\Omega}(t) = \int_{\Omega} \mathrm{d}\mathbf{x}$. Hence, in a SAB, particles almost always have a finite total length (or lifetime) $\mathcal{L}$ and eventually trap themselves in singular points in space and time. We refer to this long-time behavior as the \emph{arrested} state.
Finally, the topology of a SAB is not fixed as the generated spatial memory dynamically (and dramatically) changes the topology of the surface in a non-trivial manner. Consider a square. The topological equivalent surface of a classic square billiard is a torus (easily obtained via the process of \textit{unfolding}~\cite{masur2002rational,wright2016rational}). A self-avoiding particle generates a singular point at $t=0$, the moment it is introduced inside the square. As it begins to move ($t>0$), the singularity (now a line) extends inside the domain, resulting in surfaces with topological genus greater than 1~\cite{richens1981pseudointegrable,gutkin1984billiards}. At some point, the particle forms a new closed domain which is most likely an irrational polygon with an unidentified topological equivalent surface (see figure~\ref{fig1}c).

Importantly, the topological change results in anomalous transport of particles and memory-induced chaos: a small change in the initial condition of the particle can drastically change their trajectories as time grows. We demonstrate this in an example shown in figure~\ref{fig1}b. 
The trajectory of the two initially close particles with the same initial angle suddenly separates at the point close to their initial conditions. This bifurcation leads to significantly different trajectories, which eventually self-trap at a distance $r_f$ from each other (see video 2). 
One effective way to categorize the trajectories and demonstrate the chaos in SAB is to record the incident vector of each particle, $\mathcal{M}_{\{ p_i \}}$. Such that boundaries of the polygons with $\mathcal{N}$ edges are labelled as $A_j$, where $j \in [1,\mathcal{N}]$ and the segments of the trajectories are labelled as $l_i$, where $i \in [1, \infty)$ (see figure~\ref{fig1}c). Two initially closed particles have the same incident vector until the bifurcation moment. 
This is shown in figure~\ref{fig1}d next to the variation of the effective area $\tilde{\Omega}$ around two initially close particles (same as in figure~\ref{fig1}b) and their distance $r$. The values of $\tilde{\Omega}$ drop every time the particle traps itself in a new polygon, and clearly, $\tilde{\Omega} \rightarrow 0$ and $ r \rightarrow r_f$ as $t \rightarrow \infty$. Less evident is the probability of $r_f$ for a pair of close particles that are randomly placed in a billiard. Figure~\ref{fig1}e shows the ensemble-averaged probability density function of $r_f$ for polygons of $N \in [3,8]$. While the majority of particles stay close to each other, many end up at larger distances, sometimes more than 100 times the initial one. The probability of a larger final distance $r_f$ decays like a power-law, approximated as $\rho \sim r_f^{-p}$, where $p\approx1.4-1.7$ (with a cut-off length set by the maximum length possible in a polygon). The origin of the power-law behaviour is yet unclear to us.
%
This class of chaos observed in SAB shares similarities to the concepts of pseudo, weak or slow chaos~\cite{richens1981pseudointegrable,zaslavsky2003pseudochaos,klages2013weak,ulcigrai2021slow}, where singular topological features change the fate of initially close particles, \emph{e.g.}, in Ehrenfest billiard~\cite{ulcigrai2021slow} or billiards with barriers~\cite{eskin2003billiards,carreras2003topological}. A major difference here is that the singular features are induced by the particles themselves and hence depend on the initial particle conditions and the shape of the billiard. 

\begin{figure}[h!]
    \centering
    \includegraphics[width=0.7\textwidth]{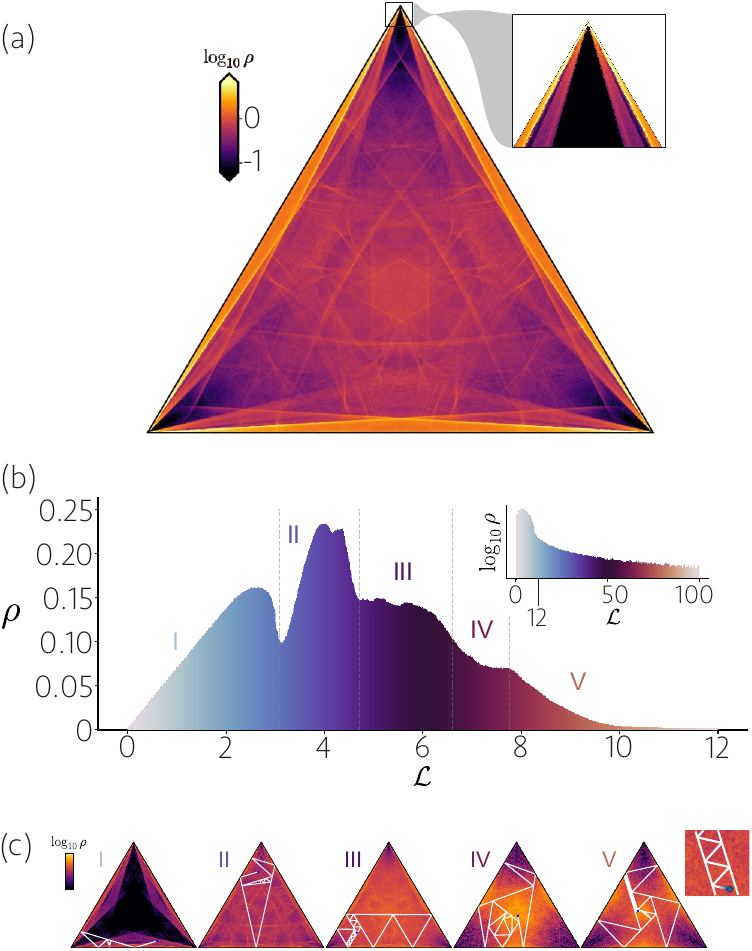}
    \caption{The \emph{arrested} state of triangular self avoiding billiard. a) Probability density ($\rho$) of final self-trapping locations. b) The total length distribution. The inset shows the semi-logarithmic version to highlight the long tail. c) Distribution of self-trapped positions in regions I to V, shown in panel b and the examples of trajectories of various lengths. The inset on the right shows a magnified view of the zigzag motion in region V.}
    \label{fig2}
\end{figure}

Given the chaotic and self-trapping nature of SAB, a natural question arises: where in space is a particle likely to become trapped? And how does this likeliness change as the (initial) geometry of the billiard change? To answer these questions, we study self-avoiding rational polygonal billiards with different numbers of edges $\mathcal{N} \in [3,\infty)$.
To this end, we perform computer simulations of $10^8$ particles with random initial position vector $(\mathbf{x}_0, \phi_0)$, where $\mathbf{x}_0$ is the position vector inside $\Omega$ and $\phi_0$ is the initial angle (see appendix~\ref{mix} for mixing and illumination tests). The particles do not interact; hence the present results are all for a single-body system (see appendix~\ref{implem} for the detail of numerical implementation). 
%
%
Figure 2a shows the probability density function of self-trapped locations $\mathbf{x}_f$ when $t \rightarrow \infty$ for the triangular billiard ($\mathcal{N}=3$). The chaotic properties of SAB results in highly complex and rich patterns. This is associated with sets (modes) of trajectories, some short-lived and some extremely long-lived. This can be seen in the distribution of the total length of the trajectories $\mathcal{L}$, shown in figure 2b (see appendix~\ref{segm} for the statistics of the line segments). For simplicity, we analyze this highly intermittent distribution in 5 different regions of total length (an alternative could be to look at sets of particles with similar incident vectors, $\mathcal{M}$). Region I belongs to short-lived particles $0<\mathcal{L} \lesssim 3$ where the particle self-traps quickly after the movement begins. The majority of these particles trap near the triangular edges. 
The distribution in region II is significantly different. Particles in this region move for longer distances and form complex structures inside the billiard. These structures suggest the presence of multiple modes of trajectories. 
Regions I and II include about 60\% (29.4\% and 29.3\%, respectively) of all particles. The rest are particles with a higher lifetime, featuring a heavy-tailed distribution (see the inset in fig~\ref{fig2}b). Particles in these regions generally don't end up near the vertices and \emph{efficiently} use the available space without self-trapping. The rare cases (extreme events) of ultra-long trajectories occur in region V. The long lifetime of these particles is a result of a zigzag motion between two almost parallel lines which were previously formed by the particle. Some of these trajectories are 20 times longer than the average trajectory length. However, the probability of their formation is less than 0.03\%.

\begin{figure}[ht!]
    \centering
    \includegraphics[width=0.9\textwidth]{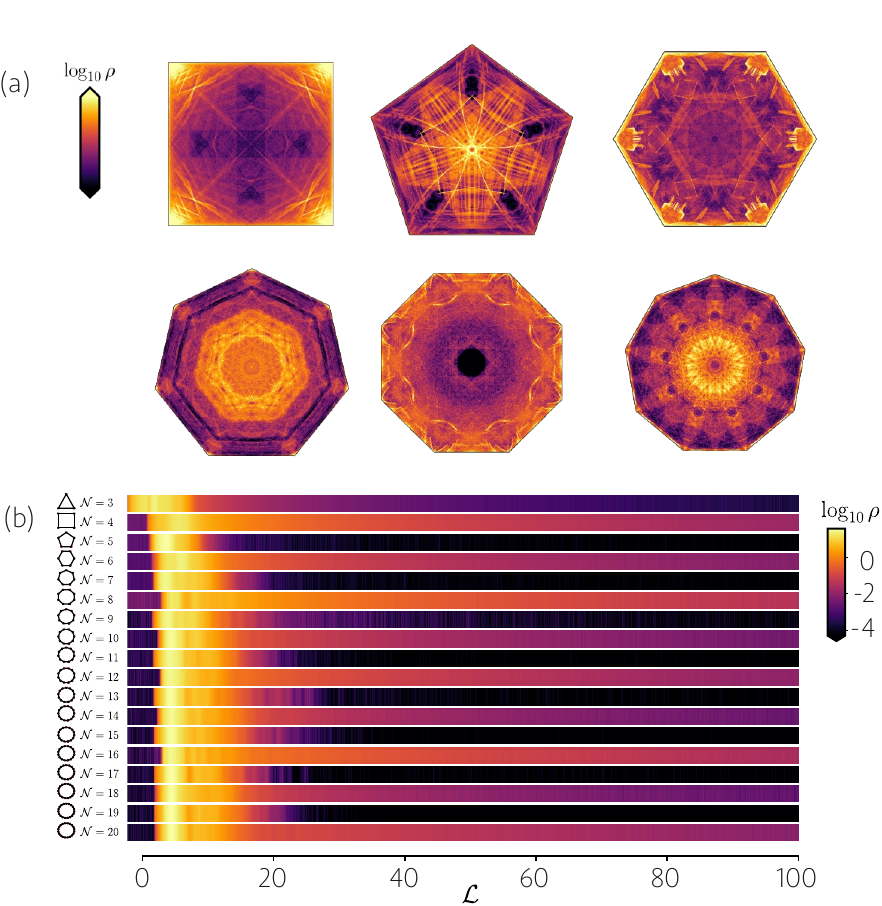}
    \caption{
    The \emph{arrested} state of polygon billiards. a) Probability density ($\rho$) of final self-trapping locations in a square, pentagon, hexagon, heptagon, octagon, and nonagon. b) The logarithmic total length distribution for regular polygons of different shapes.}
    \label{fig3}
\end{figure}

The spatial distribution of the self-trapping positions highly depends on the initial shape of the billiard. Figure~\ref{fig3}a shows the \emph{arrested} states for various regular polygons, from a square to a nonagon, where the polygons are constructed by choosing $\mathcal{N}$ equidistributed points on a unit circle. The total length distribution of these geometries and also polygons with a higher number of vertices are shown in figure~\ref{fig3}b. The $\mathcal{N}$-fold symmetric final patterns clearly vary with the geometry of the billiard. But a few features seem to be universal. Particles in even polygons tend to trap more near the vertices and also have a higher chance of a long lifetime ($\mathcal{L}$), since the polygon itself features parallel walls, allowing for zigzag bounces. In contrast, the  self-trapping probability in the centre is higher for odd polygons, and the probability of a long lifetime is low. Notably, the triangle is the only polygon for which highly likely arbitrarily small orbits are possible since it has interior angles smaller than $\pi / 2$.



The results presented here illustrate the complex nature of dynamical systems with spatial memory. In contrast to previous studies on active particles with memory (\emph{e.g.}, those used in~\cite{schulz2005feedback, kranz2019trail, gelimson2016multicellular, kranz2016effective, sengupta2009dynamics, daftari2022self}), the current deterministic framework, based on mathematical billiards, employs extremely simple microscopic rules without noise or particle interaction. Yet, complex patterns and anomalous transport emerge due to memory-induced topological changes.

We found that ballistic particles with spatial memory self-trap and exhibit topology-induced chaos. These dynamical characteristics make it non-trivial to predict the long-time asymptotic behavior of the system. Nonetheless, this limit can be accessed through numerical simulations. As a dynamic system, a Self-Avoiding Billiard (SAB) fundamentally differs from classic billiards because the surface on which particles flow evolves over time, and the shape of the polygon almost always morphs into an irrational one, which is considerably more challenging to treat mathematically. Nevertheless, the initial shape of the polygon governs the final \emph{arrested} state, as demonstrated in figure~\ref{fig3}.

There are several immediate opportunities to extend the findings of this work. Billiards of different geometries in elliptic or hyperbolic systems (\emph{e.g.}, stadium or Sinai billiards), exhibit fundamentally different ergodic and chaotic behavior. Combining spatial memory with such billiards complements the present study. 
Additionally, in biological or physicochemical systems, spatial memory often dissipates over time as chemical trails diffuse. This introduces another timescale $t_m$. The ratio of the particle's convective timescale to the fading timescale (known as the Péclet number in hydrodynamics) governs the dynamics of the system. In the current study, $t_m \rightarrow \infty$, indicating permanent memory. However, for finite values of $t_m$, one may observe both self-trapping and cage breaking, leading to different long-time behavior. 
Moreover, the particle's reaction to its memory and the boundaries represents another control parameter. Here, we consider the simplest form of elastic collision for all interactions. Inelastic collisions~\cite{spagnolie2017microorganism, eyles2011interference, thery2021rebound, besemer2022glory} or probabilistic collisions can significantly alter the system's dynamics. The study of many-body SAB is also of particular interest since particle interactions can take various forms, including reciprocal and non-reciprocal interactions~\cite{durve2018active, meredith2020predator, kreienkamp2022clustering, osat2023non}. Furthermore, in terms of practical applications, given the simplicity of the rules employed in this study, spatial memory could be utilized to optimize autonomous robotic systems~\cite{burgard2000collaborative, garcia2021intrinsic} and active matter~\cite{nakayama2023tunable, hokmabad2022chemotactic}, especially when combined with learning techniques~\cite{kaelbling2020foundation, ecoffet2021first}.

\pagebreak
\clearpage
\section*{Supplementary Materials}

\subsection{Numerical implementation}
\label{implem}
The algorithm below is the pseudocode used to perform the SAB simulations. $\Omega$ is the interior of the billiard table, with the boundary given by $\partial \Omega$.
$\textbf{p}$ represents the particle. It has a position and velocity which we choose uniformly random at the beginning of a simulation. $W$ is a wall object, which is a line segment with a normal vector $\hat{n}$. $M$ is the matrix used to calculate the time till collision $t$ with a certain wall, and the parametric position on the wall $s$. The numerical code is implemented in Julia~\cite{bezanson2017julia} (also see~\cite{Datseris2017}).

\begin{algorithm}
\caption*{Self avoiding billiard algorithm}\label{alg:cap}
\begin{algorithmic}

\Function{Collision}{$\textbf{p},\ \text{W}$}
\State $\overrightarrow{sp} \gets \text{W.sp}$
\State $\overrightarrow{ep} \gets \text{W.ep}$
\State $M \gets \begin{pmatrix}
\overrightarrow{ep}_x - \overrightarrow{sp}_x & -\textbf{p}\text{.vel}_x\\
\overrightarrow{ep}_y - \overrightarrow{sp}_y & -\textbf{p}\text{.vel}_y
\end{pmatrix}$
\State $\begin{pmatrix}
    s\\ t
\end{pmatrix} \gets M^{-1} (\textbf{p}\text{.pos} - \overrightarrow{sp})$

\If{$0 < s < 1$} \Comment{Collision must be on wall}
    \State \Return $t,\ \textbf{p}\text{.pos} + t\cdot \textbf{p}\text{.vel}$
\Else
    \State \Return $\infty,\ \vec{0}$ \Comment{No collision found}
\EndIf
\EndFunction
\\
\Function{Reflect}{$\textbf{p}$, $\partial \Omega$}
\State $t_{min} \gets \infty$
\State $\vec{x}_{min} \gets \vec{0}$
\For{$W \in \partial \Omega$}
    \State $t,\ \vec{x}_{cp} \gets$ \Call{Collision}{$\textbf{p},\ \text{W}$}
    \If{$t < t_{min}$}
        \State $t_{min} \gets t$
        \State $\vec{x}_{min} \gets \vec{x}_{cp}$
    \EndIf
\EndFor
\State $\hat{n} \gets \text{W.normal}$
\State $\textbf{p}\text{.vel} \gets \textbf{p}\text{.vel} - 2 (\hat{n} \cdot \textbf{p}\text{.vel}) \hat{n}$
\State $\textbf{p}\text{.pos} \gets \vec{x}_{min}$
\State \Return $t_{min}$,\ $\vec{x}_{min}$
\EndFunction
\\
\State $\vec{x}_{i} = \begin{pmatrix}
    \cos \frac{i}{2\pi n}\\
    \sin \frac{i}{2\pi n}
\end{pmatrix}$ \Comment{Build polygonal table}
\State $\partial \Omega \gets \{(\vec{x}_i, \vec{x}_{i+1})\ |\ \forall i \in \{1, \dots, n\}\}$ 
\State $\textbf{p}\text{.pos} \in \Omega$ \Comment{Pick random initial conditions}
\State $\textbf{p}\text{.vel} \in S^1$

\While{$t > \epsilon$}
\State $\vec{x}_{sp} \gets \textbf{p}\text{.pos}$
\State $t,\ \vec{x}_{ep} \gets $\Call{Reflect}{$\textbf{p}$, $\partial \Omega$}
\State $\partial \Omega \gets \partial \Omega \cup \{(\vec{x}_{sp},\ \vec{x}_{ep})\}$ \Comment{Add wall}
\EndWhile
\end{algorithmic}
\end{algorithm}

\subsection{Statistics of segments and angles}
\label{segm}
\begin{figure}[H]
    \centering
    \includegraphics[width=0.8\textwidth]{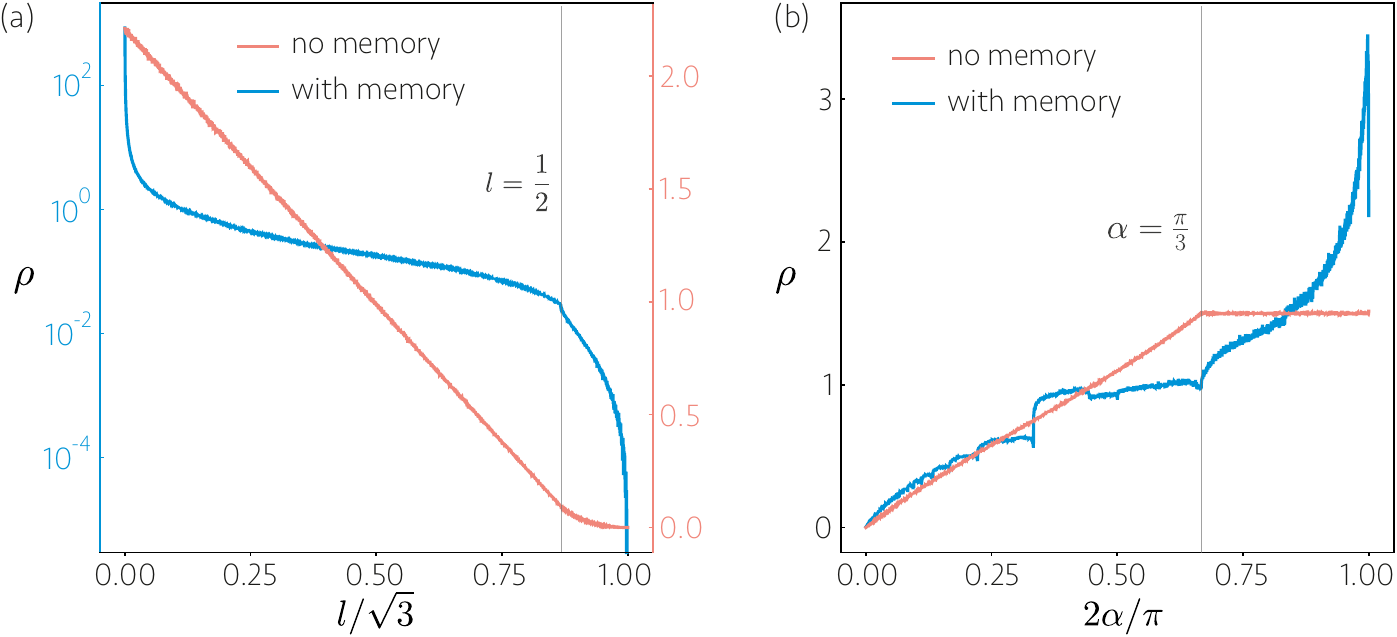}
    \caption{Distribution of a) segment's length (note the double y-axis) and b) incident angle in triangular classic (no memory) and self-avoiding (with memory) billiards. The black vertical lines show the length and angle of the triangle.}
    \label{segs}
\end{figure}

The introduction of self-avoiding memory significantly changes the statistics of line segments (each individual collision) and their incident angles (see figure~\ref{segs}). Memory leads to a large probability of small segments due to local collisions around the self-trapping point and significantly reduces the change of any segment in the order of the polygon's length. The distribution of incident angles is also peculiar with various jumps. Although a The probability of angles close to $\pi/2$ is much higher in SAB, due to the formation of non-triangular billiards during self-entrapment. Note that, although the probability of angles in a classic billiard (no memory) is attainable via relatively easy geometrical consideration, we yet to find a good geometrical description for SAB.

\subsection{Mixing and illumination}
\label{mix}
\begin{figure}[H]
    \centering
    \includegraphics[width=1\textwidth]{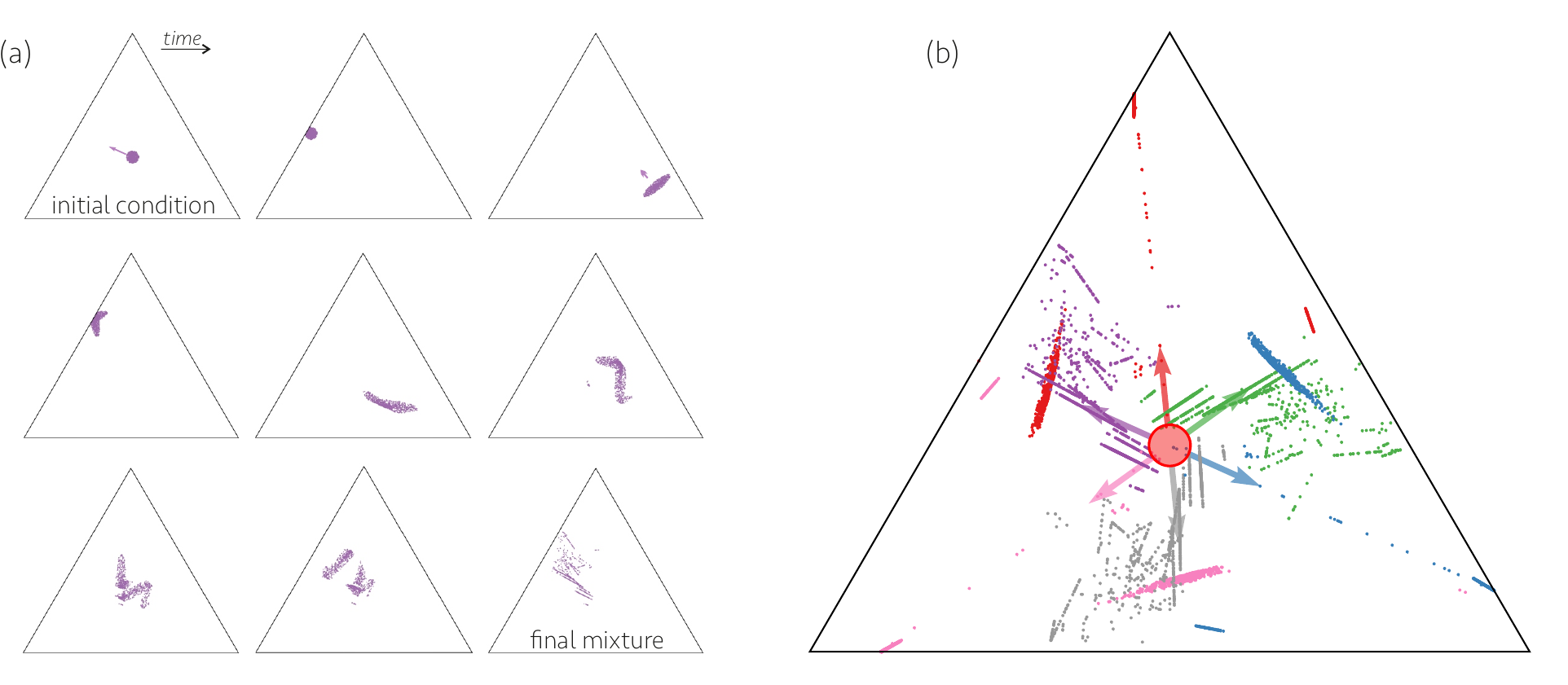}
    \caption{Example of anomalous chaotic mixing in a triangular self-avoiding billiard. a) Particles at the centre of the triangle flow in the same direction, shown by the vector. The self-avoiding results in a final state where particles are weakly mixed (see video 3). b) The final state depends on the initial condition. Here we show how changing the direction of the initial angle changes the final mixtures, shown by different colors.}
    \label{mixing}
\end{figure}

The self-trapping and finite lifetime of the particles in SAB leads to inherently non-ergodic dynamics. Consequently, the system shows an anomalous weakly chaotic mixing characteristic as particles reach the \emph{arrested} state. A visual representation of such a behaviour can be seen in figure~\ref{mixing} (see video 3). A collection of initially closely-distanced particles (in the circle in the centre of the triangle) flow and eventually self-trap in a set of locations. The final mixing state depends on the initial condition (and the shape of the polygon), here shown by changing the initial angle of motion (shown by 6 different vectors in figure~\ref{mixing}b).

\begin{figure}[H]
    \centering
    \includegraphics[width=0.85\textwidth]{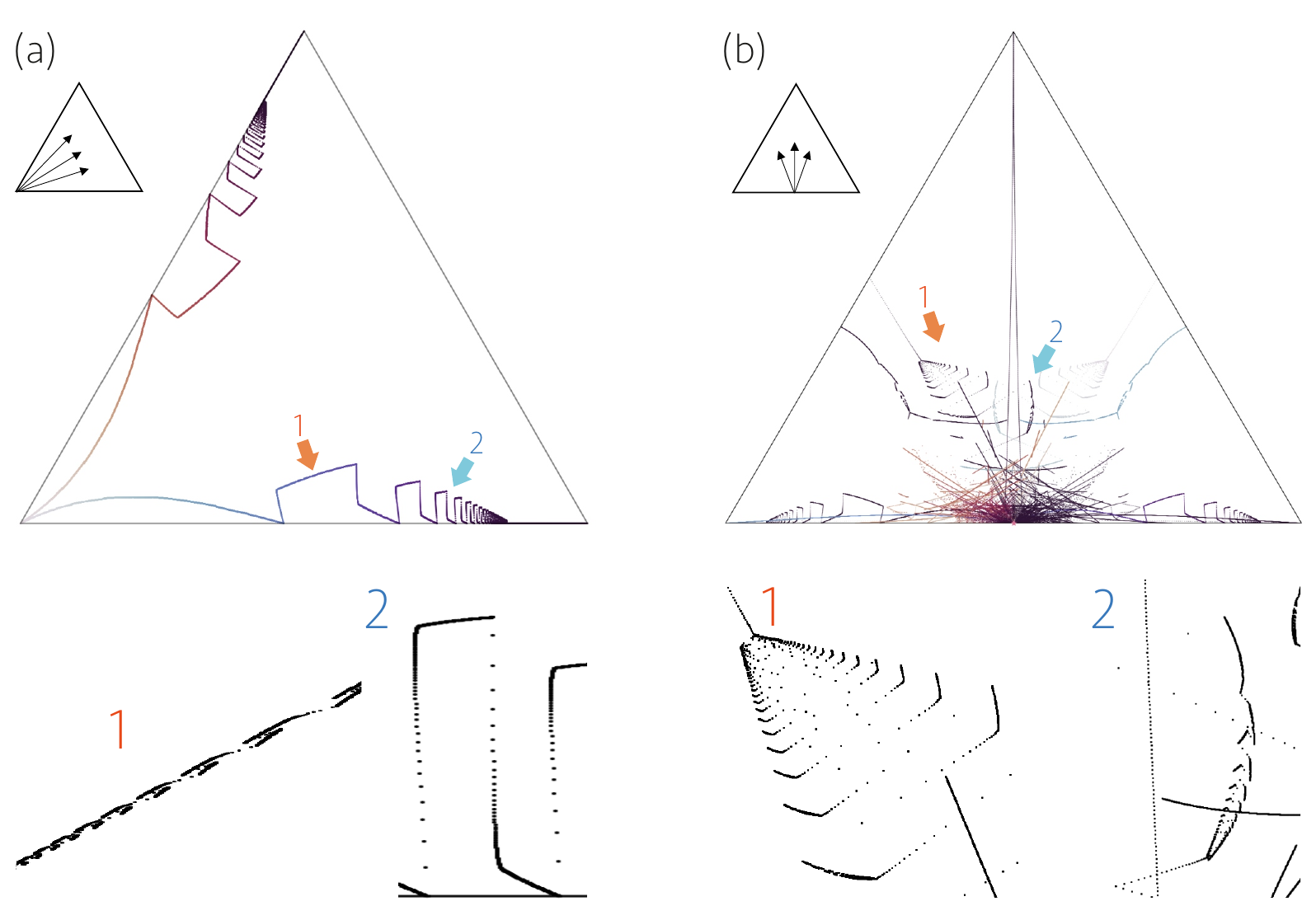}
    \caption{The final state of two illumination tests when the light source (point) is located at a) left vortex and b) between left and right vertices. The insets in the bottom show the magnified view of the small structures (see video 4).}
    \label{illum}
\end{figure}

The same features (self-trapping and chaos), similarly, result in complex entrapment of light in illumination problems~\cite{klee1969every,tokarsky1995polygonal,lelievre2016everything,wolecki2019illumination,numberphile}. In an illumination problem, a light source is placed a particular location and one observes which part of the room is illuminated and which part remains dark. To this end, the particle trajectories and ray traces are equivalent. Figure~\ref{illum} exhibit two examples of illumination in a triangular SAB (see videos 4 and 5). Placing the \say{light source} at a vertex of the triangle results in a complex shape of a trapped positions (figure~\ref{illum}a). Hence, in log-term, everywhere inside a triangular SAB is dark, except at these positions. Such final illumination is even more complex with severals small structures when the light source is placed in the centre of the triangle edge (figure~\ref{illum}b).

\section*{Supplementary Videos}
\textbf{Video 1:} Examples of classic and self avoiding billiards for a particle with the same initial conditions.\\
\textbf{Video 2:} Topology-induced chaos in a self-avoiding billiard: two initially close particles separate from each other and self-trap themselves at a larger distance.\\
\textbf{Video 3:} Anomalous mixing in a self-avoiding billiard. A \say{droplet} of particles reach an \emph{arrested} state after interacting with self-created spatial memory.\\
\textbf{Video 4:} The illumination tests in a triangular billiard tests where the light source is placed at a vertex. 4 trajectories are also shown for clarity. \\
\textbf{Video 5:} The illumination tests in a triangular billiard tests where the light source is placed between two vertices. 4 trajectories are also shown for clarity.

\section*{Acknowledgements}
The authors would like to thank Alvaro Marin, and Clélia De Mulatier for insightful discussions.

\printbibliography

\end{document}